\documentclass[prb,aps,twocolumn]{revtex4}

\usepackage{graphicx}

\begin{document}

\title{Linear scaling calculation of maximally-localized Wannier functions 
with atomic basis set}

\author{H. J. Xiang}
\affiliation{Hefei National Laboratory for Physical Sciences at
  Microscale,
  University of Science and Technology of
  China, Hefei, Anhui 230026, People's Republic of China}

\author{Zhenyu Li}
\affiliation{Hefei National Laboratory for Physical Sciences at
  Microscale,
  University of Science and Technology of
  China, Hefei, Anhui 230026, People's Republic of China}

\author{W. Z. Liang}
\affiliation{Hefei National Laboratory for Physical Sciences at
  Microscale,
  University of Science and Technology of
  China, Hefei, Anhui 230026, People's Republic of China}

\author{Jinlong Yang}
\thanks{Corresponding author. E-mail: jlyang@ustc.edu.cn}

\affiliation{Hefei National Laboratory for Physical Sciences at
  Microscale,
  University of Science and Technology of
  China, Hefei, Anhui 230026, People's Republic of China}

\author{J. G. Hou}
\affiliation{Hefei National Laboratory for Physical Sciences at
  Microscale,
  University of Science and Technology of
  China, Hefei, Anhui 230026, People's Republic of China}

\author{Qingshi Zhu}
\affiliation{Hefei National Laboratory for Physical Sciences at
  Microscale,
  University of Science and Technology of
  China, Hefei, Anhui 230026, People's Republic of China}

\date{\today}

\begin{abstract}
  We have developed a linear scaling algorithm for calculating
  maximally-localized Wannier functions (MLWFs) using atomic orbital basis.
  An O(N) ground state calculation is carried out to get the
  density matrix (DM).
  Through a projection of the DM onto atomic orbitals and a subsequent O(N)
  orthogonalization, we obtain initial orthogonal localized
  orbitals. These orbitals can be maximally localized
  in linear scaling by simple Jacobi sweeps. Our O(N)
  method is validated by applying it to water molecule and wurtzite
  ZnO. The linear scaling behavior of the new method is demonstrated by
  computing the MLWFs of boron nitride nanotubes.
\end{abstract}


\maketitle

\section{Introduction}
\label{intro}
Wannier function\cite{WF} is a powerful tool in the study of the
chemical bonding, dielectric properties, excited
electronic states, electron transport, and many body correlations in materials.
In particular, the modern theory of bulk polarization
relates the vector sum of the centers of the Wannier functions to the
macroscopic polarization of a crystalline insulator.\cite{mod_polarization}
However, the intrinsic nonuniqueness in the
Wannier function definition, and the difficulty in defining
their centers within a periodic cell calculation, limited
their practical use. Fortunately,
an elegant method has been recently
proposed by Marzari and Vanderbilt to obtain a unique set of
maximally-localized Wannier functions (MLWFs).\cite{MLWF}
By transforming the occupied electronic manifold into a set of MLWFs,
it becomes possible to obtain an enhanced understanding of chemical
bonding properties and electric polarization via an analysis of the MLWFs.
Beside the above points, the MLWFs are now also
being used as a very accurate minimal basis for a variety of algorithmic or
theoretical developments, with recent applications ranging from
linear-scaling approaches\cite{MLWF_QMC} to the construction of effective
Hamiltonians for the study of ballistic transport,\cite{MLWF_trans}
strongly-correlated electrons,\cite{MLWF_strong} self-interaction
corrections, metal-insulator transitions,\cite{MLWF_MIT} and photonic
lattices.\cite{MLWF_photon}

In the seminal work of Marzari and Vanderbilt, first a ground state
calculation was carried out to obtain the occupied delocalized canonical
orbitals, then a sequence of unitary transformations were performed to
obtain MLWFs which minimize the spread function.\cite{MLWF}
Using the exponential representation for the unitary transformation,
Berghold {\it et al.}\cite{MLWF_gamma1} derived an iterative scheme  
to obtain MLWFs in large supercells of
arbitrary symmetry. Also a simple Jacobi orbital
rotation scheme was found to be remarkably
efficient.\cite{MLWF_gamma1}
A simultaneous diagonalization algorithm similar to the Jacobi
diagonalization method, was used by Gygi {\it et al.} to 
compute MLWFs.\cite{MLWF_diag} 
Zicovich-Wilson {\it et al.} proposed a Wannier-Boys scheme  to obtain
well localized Wannier functions in linear combination of atomic
orbital periodic calculations. \cite{crystal} 
However, all methods mentioned above  for
calculating MLWFs are nearly O(N$^3$) scaling (N is the number of electrons),
which prohibits their applications to large systems containing
hundreds or thousands of atoms. The unfavorable scaling
comes from two steps in these methods: The conventional
methods for getting ground state wavefunctions is
O(N$^3$) or O(N$^{2} \ln$N), and the localization step in the above
localization algorithms is also O(N$^3$). Usually, the traditional
ground state calculation will cost more than the localization step.  
However, for large systems the computing amount of the localization
step  is also time-consuming. 

In this work, we propose a simple order-N algorithm for effectively
calculating MLWFs.
The demanding ground state calculation is circumvented by using O(N)
density matrix purification methods. After adopting O(N) method for
the ground state calculation, the conventional O(N$^3$) localization
step will become time-dominant 
for large systems. To obtain MLWFs in linear scaling, we first get
initial localized orbitals from the density matrix,
then an O(N) localization method which uses the Jacobi rotation scheme
is utilized to maximally localize the orbitals.   
The linear scaling behavior of
the new method is demonstrated by computing the MLWFs of boron nitride (BN)
nanotubes. 

This paper is organized as follows: In Sec.~\ref{theory}, we
present our new O(N) method for calculating MLWFs.
In Sec.~\ref{impl},
we describe the details of the implementation and perform
some test calculations to illustrate the rightness, robustness,
and linear-scaling behavior of our methods. 
We discuss some possible extensions and generalizations of our
method in Sec.~\ref{dis}. 
Finally, our concluding remarks are given in
Sec.~\ref{con}.

\section{Theory}
\label{theory}
\subsection{Maximally-Localized Wannier Functions}
The Wannier functions are defined in terms of a unitary transformation
of the occupied Bloch orbitals. 
However, they are not uniquely defined, due to the arbitrary freedom in
the phases of the Bloch orbitals. Marzari and Vanderbilt \cite{MLWF} resolve this 
indeterminacy by minimizing  the total spread function of 
the Wannier functions $w_{n}(\mathbf{r})$ 
\begin{equation}
S=\sum_{n} (\langle r^{2} \rangle_{n} - \langle r \rangle_{n}^{2} ),
\end{equation}
where $\langle r \rangle_{n} = \langle w_{n} | r | w_{n} \rangle$, 
and $\langle r^{2} \rangle_{n} = \langle w_{n} | r^{2} | w_{n} \rangle$.

Here since we aim at large systems, the $\Gamma$-point-only sampling 
of the Brillouin zone (BZ) is used throughout this work. 
The method of calculating MLWFs for supercells of general symmetry is 
proposed by Silvestrelli {\it et al.}.\cite{MLWF_gamma2}
For the sake of simplicity, considering the case of a simple-cubic supercell of
side $L$, it can be proved that minimizing the total spread $S$ is
equivalent to the problem of maximizing the functional
\begin{equation}
\Omega=\sum_n (|X_{nn}|^2+|Y_{nn}|^2+|Z_{nn}|^2),
\end{equation}
where $X_{mn}=\langle w_{m} |e^{-i\frac{2\pi}{L} x}|w_{n}\rangle$ and 
similar definitions for $Y_{mn}$ and $Z_{mn}$ apply. 
The coordinate
$x_n$ of the $n$th Wannier-function center (WFC) is computed
using the formula
\begin{equation}
  x_n = - \frac{L}{2\pi}\mathrm{Im} \mathrm{ln} \langle w_{n} |
  e^{-i\frac{2\pi}{L} x} | w_{n} \rangle, 
\end{equation}
with similar definitions for $y_n$ and $z_n$.

\subsection{Our O(N) method for calculating MLWFs}
Our new O(N) method consists of four O(N) steps: first we obtain the density
matrix, secondly we find out a set of linear independent nonorthogonal
orbitals which span the occupied manifold, thirdly, 
a modified L\"owdin orthogonalization is used to orthogonalize these
nonorthogonal orbitals, finally, the Jacobi rotation scheme  is utilized to
maximally localize the orbitals.

In principles, any localized orbitals or density
matrix based linear scaling methods
can be used to obtain initial localized orbitals in linear
scaling.\cite{DM_rev}
Here we use the O(N)
trace-correcting density matrix purification (TC2)\cite{DM11}
method to get the density matrix since it is very simple, robust, and
efficient. 
The use of some other linear scaling methods based on
localized orbitals will be discussed in Sec.~\ref{dis}.
Here, the essence of the TC2 method is briefly outlined. 
In the begining, the Hamiltonian $H$ represented in the
orthogonal basis 
is normalized to an initial matrix $\rho_0$ with all its eigenvalues 
mapped onto [0,1]: 
$\rho_0 = (\epsilon_{max} I - H) / ( \epsilon_{max} - \epsilon_{min}
)$, where $\epsilon_{min}$ and $\epsilon_{max}$ are lowest and highest
eigenvalues of $H$, respectively. Then we correct the trace of the
density matrix while purifying it using the following iteration: 
\begin{equation}
\rho_{n+1}(\rho_{n}) =
\left\{\begin{array}{ll}
\rho_{n}  ^2, &  Tr(\rho_{n} ) \geq N_{e}/2 \\
2\rho_{n} - \rho_{n}^2, & Tr(\rho_{n}) <
N_{e}/2,
\end{array} \right.
\end{equation}
where $N_{e}$ is the total number of electrons in a close-shell system.

Given an atomic orbital, one can
project out its occupied component using the density matrix operator $\hat{P}$:
\begin{equation}
   \Phi_{\alpha}= |\hat{P} \phi_{\alpha}\rangle=\sum_\beta
   (\mathbf{P}\mathbf{S})_{\beta,\alpha} |\phi_{\beta}\rangle,
\end{equation}
where $\mathbf{S}$ is the overlap matrix, $\phi_\alpha$ and $\phi_\beta$ denote the
atomic basis orbitals, $\mathbf{P}$ denotes the density
matrix in the atomic orbital basis. By applying the density matrix
operator on $N_{b}$ atomic basis orbitals, we can get $N_{b}$
localized orbitals $\Phi_{\alpha}$, among which
only $N_{occ}$ (the number of occupied
states) localized orbitals are linear independent. One must select out
$N_{occ}$  linear independent localized orbitals among these
localized orbitals $\Phi_{\alpha}$ before performing localization
steps.
We have implemented two algorithms to achieve this goal.
One of the  algorithms is similar to that proposed by 
Maslen {\it et al.},\cite{DM_gordon} who used Cholesky decompositions
for detecting the linear 
dependence.
We note that in this method the total demanding for the construction
of all the overlap matrices is O(N) since the nonorthogonal orbitals are
localized. Furthermore, the total computing amount for performing all Cholesky
decompositions is almost the same as that for performing a sparse
Cholesky decomposition with the matrix dimension $N_{occ}$ due to the
nature of the Cholesky decomposition. In the second algorithm, we
select these atomic basis orbitals to be projected according to the physical
intuition. For example, in BN nanotubes, there are
one 2s and three 2p basis orbitals for each B and N atoms when using
pseudopotentials. Since some
electrons will transfer from B to N atoms, we can get $N_{occ}$ linear
independent localized orbitals by just projecting the density matrix on all
atomic basis orbitals of N atoms in despite of the large covalency in
these systems. For systems where the bonding properties are known, the
algorithm is found to be very efficient and the resulting orbitals are
very sparse. In cases where the second algorithm doesn't apply, we
will resort to the first algorithm.

Since MLWFs are orthogonal to each other, the $N_{occ}$ linear
independent nonorthogonal localized orbitals must be orthogonalized.
It is well known that the orthogonalized orbitals produced by the L\"owdin
orthogonalization are closest to initial nonorthogonal
orbitals in the sense of least squares. To obtain orthogonal
localized orbitals in linear scaling, 
we carry out a modified L\"owdin
orthogonalization adopted by Stephan and Drabold.\cite{orth}
In this
approach, we perform repeated first-order L\"owdin iterations
\begin{equation}
  \Phi_{\alpha}^{'}=\Phi_{\alpha}-\frac{1}{2}\sum_{\beta(\ne \alpha)}
  \Phi_{\beta} \langle\Phi_{\alpha}|\Phi_{\beta}\rangle.
\end{equation}
The functions after every orthonormalization cycle have to be
renormalized. Typically, we can obtain well orthogonal localized
orbitals in about five O(N) orthonormalization cycles. 

Now we will discuss how to maximally localize these orthogonal
orbitals to get MLWFs in linear scaling.
Our target is maximizing
$\Omega$ to obtain MLWFs in linear scaling.
The key point for the successful O(N) localization step is that
these orbitals are localized in the whole localization procedure.
We use the simple Jacobi rotation method to maximize $\Omega$ since it
doesn't require O(N$^3$) diagonalization in contrast to the unitary
transformation method.\cite{MLWF_gamma1} 
This method is a traditional method in quantum
chemistry for computing localized molecular orbitals first introduced
by Edmiston and Ruedenberg.\cite{ER1} The
basic idea of the method is to tackle the problem of maximizing
$\Omega$ by carrying out several Jacobi sweeps. In a traditional
Jacobi sweep, we perform $N_{occ}(N_{occ}-1)/2$ consecutive two-by-two
rotations among all pairs of orbitals. The elementary step consists
of a plane rotation where two orbitals are rotated through an angle
and all other orbitals are fixed. 
In our O(N) method, due to the localization of
the orbitals, the computing amount of an orbital ratation is O(N).
Moreover, each orbital overlaps with only O(1) orbitals, thus the 
number of the Jacobi rotations in a Jacobi sweep is of order N. 
Assuming the number of the Jacobi sweeps doesn't change (It is
really the case for systems with similar characters), the total
computing amount is O(N).

\begin{figure}[!hbp]
  \includegraphics[width=6.5cm]{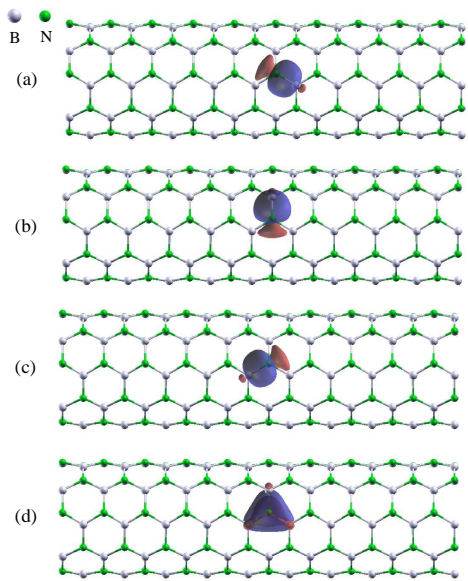}
  \caption{(Color online)
    (a), (b), (c), and (d) show the four MLWFs around a nitrogen atom of BN(5,5)
    nanotubes. These MLWFs are calculated using the supercell containing 200
    atoms with the DZP basis set.}
  \label{fig1}
\end{figure}

\begin{figure}[!hbp]
  \includegraphics[width=6.5cm]{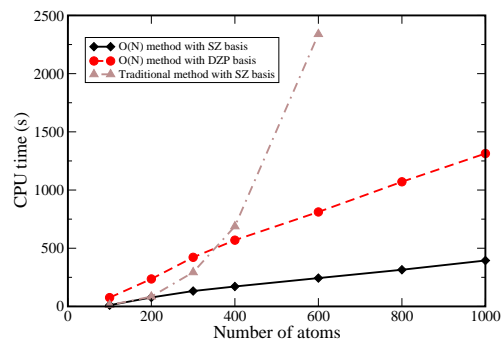}
  \caption{(Color online)
    Total CPU time for calculating MLWFs of BN(5,5) nanotubes
    using the linear scaling method or the traditional Jacobi rotation
    method which doesn't take advantage of the localization property of the
    orbitals.  In case of the new O(N) method, both SZ and DZP basis sets
    are used. The calculations using the traditional method are performed
    using the SZ basis set. All calculations were carried out on a 1.5 GHz
    Itanium 2 CPU workstation running RedHat Linux Advanced Server V2.1.}
  \label{fig2}
\end{figure}

\section{Implementation and Results}
\label{impl}
\subsection{Implementation}
Our newly developed method has been implemented in SIESTA,\cite{siesta} a standard
Kohn-Sham density-functional program using norm-conserving
pseudopotentials and numerical atomic orbitals as basis sets. 
In SIESTA, periodic boundary conditions are
employed to simulate both isolated and periodic systems.
The details about the implementation of the TC2 method
can be found in Ref. [18]. 

\subsection{Validity and performance of the method}
All our calculations reported in this work are done in the
local density approximation (LDA).\cite{LDA} Unless otherwise stated, 
the double-$\zeta$ plus polarization functions (DZP) basis set is used
in the calculations.
First we calculate the MLWFs (i.e., Boys orbitals\cite{Boys}) of a water
molecule. It is well known that there are four MLWFs for a H$_2$O
molecule: two covalent O-H $\sigma$ bonds and two lone-pair orbitals.
The distance between the centroids of these four MLWFs and the
position of the oxygen ion is 0.52, 0.52, 0.30, and 0.30 \AA\ respectively.
The results agree well with those reported by Berghold 
{\it et al.}.\cite{MLWF_gamma1}  
As a second check of the validity of our method, we calculate the 
piezoelectric constant of bulk ZnO.
Both the Berry phase method\cite{mod_polarization} and our new O(N) method are
used to calculate piezoelectric constant $e_{33}$ of bulk wurtzite
ZnO. In our O(N) calculation, we use a $6\times 6 \times 2$ ZnO supercell since
we use the $\Gamma$-only sampling.
The results from these two methods agree very well:
The computed values are $1.29$ and $1.30$ C/m$^2$, respectively. 
And both results accord with others' result \cite{DFPT_Vanderbilt}
($1.28$ C/m$^2$) computed through the density
functional perturbation theory. \cite{DFPT_Vanderbilt}

Then we test our method by applying it to calculate the MLWFs of
BN(5,5) armchair nanotubes.
Fig.~\ref{fig1} shows four MLWFs of BN(5,5) nanotubes computed using 
the supercell  containing 200 atoms.
We can clearly see that
the three MLWFs in Fig.~\ref{fig1}(a)-(c) are B-N $\sigma$
bonds. Among these MLWFs, the MLWF in Fig.~\ref{fig1}(a) has
exactly the same character as that shown in Fig.~\ref{fig1}(c) due to
the mirror plane symmetry in armchair BN nanotubes. 
The other MLWF in Fig.~\ref{fig1}(d) is a $\pi$ orbital,
which almost centers at a nitrogen atom.
It is interesting
that in BN nanotubes MLWFs preserve $\sigma-\pi$ separation.
We will return to this point later. 
To see the efficiency of our new O(N) method, we perform a series of
calculations using the linear scaling method or the traditional Jacobi rotation
method  which doesn't take advantage of the localization property of
the orbitals.   
Two different basis sets (single-$\zeta$ (SZ), DZP) are employed in the calculations using
the new O(N) method, and only the SZ basis set is used for the
traditional method.
The CPU time is shown in Fig.~\ref{fig2}.
We clearly see the perfect linear scaling behavior of our new method.
And the traditional method displays a nearly O(N$^3$) scaling as
expected. The computing saving of our method with respect to the traditional
method is dramatically large, especially when the size of systems
exceeds 400 atoms. 
In addition, we note that the ratio between the time for calculating MLWFs using
DZP basis and that using SZ basis is smaller than the case for the
ground state calculation.\cite{SU_TC2}

\section{Discussion}
\label{dis}
We have also implemented the method for obtaining MLWFs from the localized
orbitals produced by the O(N) Mauri-Ordej\'on (MO) \cite{OM1,OM2} or KMG\cite{KMG}
methods. In case of the MO method, the number of localized orbitals is
equal to the number of occupied states, and all localized orbitals
produced by the O(N) MO method are linear independent. Thus the
projection step is unnecessary. However, in the KMG energy functional,
the number of localized orbitals is larger than the number of
occupied states.  
In this case, we first get the density matrix 
using Equation (87) in the paper written by Soler {et al.}.\cite{siesta} 
Once the density matrix is available, the steps to get MLWFs are the
same as those described in Sec.~\ref{theory}.

In our implementation, numerical localized atomic orbitals are used as basis sets. 
Other localized basis such as Gaussian orbitals or real space methods\cite{fd}
can also be used. 
However, the use of non-local plane-wave basis which was commonly adopted in
previous calculations of MLWFs will not result in linear scaling
behavior.

Our previous discussions mainly focus on MLWFs in periodic systems. 
We should note that our method could also be adopted to obtain
Boys\cite{Boys} localized orbitals of isolated molecular systems with
or without using periodic boundary conditions. In this case, one can
directly minimize the spread function instead maximize the functional
$\Omega$. Besides Boys orbitals, Edmiston-Ruedenberg (ER)\cite{ER1,ER2} and
Pipek-Mezey (PM)\cite{PM} localized orbitals are also popular among
chemists. The benefit of these localized orbitals  is that, unlike Boys
orbitals, ER and PM orbitals always preserve $\sigma-\pi$ separation.
Our method can be used to get PM orbitals in linear scaling.
However, due to the long range character of the operator $1/\mathbf{r}$, our
method is unable to reduce significantly the computing amount in the
calculation of ER orbitals. We notice that an efficient method which
reduces the scaling from O(N$^5$) to O(N$^2$)-O(N$^3$) has been
proposed by Subotnik {\it et al.}.\cite{homogeneous}

Although we only discuss spin-restricted systems up to now, combined
with the spin-unrestricted linear scaling electronic structure
theory,\cite{SU_TC2} 
our O(N) method can be straightforwardly applied to insulating
magnetic systems. This method can be used to study the large
multiferroic (simultaneously (ferro)magnetic and ferroelectric)
materials. 

\section{Conclusions}
\label{con}
To summarize, a linear scaling algorithm for calculating MLWFs
has been proposed for the first time. The numerical atomic orbital
basis instead of commonly adopted plane wave sets is used. First we
perform a linear scaling ground state calculation using the TC2
purification method. From the density matrix, we get the initial
non-orthogonal localized orbitals. Through a modified L\"owdin
orthogonalization, we obtain the initial orthogonal localized
orbitals. Due to the localization property of these initial orbitals,
the computing requirement of the subsequent Jacobi sweeps for getting
MLWFs is also linear scaling. Our results for water molecule and bulk
wurtzite ZnO agree well with others' results. The O(N) behavior of the
proposed method is clearly demonstrated by computing the MLWFs of BN 
nanotubes. Our O(N) method provides a very efficient way for obtaining
MLWFs which have many possible applications. 

This work is partially supported by the National Natural Science Foundation of China
(50121202, 20533030, 10474087), by the USTC-HP HPC project, and by the
SCCAS and Shanghai Supercomputer Center.


\end{document}